\documentclass{jpsj2}
\usepackage{bm}
\title{
  Finite-Temperature Transition in the Spin-Dimer Antiferromagnet BaCuSi$_2$O$_6$
}

\author{
  \textsc{Yoshitomo Kamiya},
  \textsc{Naoki Kawashima}
  and 
  \textsc{Cristian D. Batista}$^{1}$
}

\inst{
  Institute for Solid State Physics, University of Tokyo, Kashiwa,
  Chiba 227-8581, Japan \\
  $^{1}$Theoretical Division, Los Alamos National Laboratory, Los Alamos, New
  Mexico 87545, USA
}

\abst{
  We consider a classical XY-like Hamiltonian on a body-centered tetragonal
  lattice, focusing on the role of interlayer frustration. A three-dimensional
  (3D) ordered phase is realized via thermal fluctuations, breaking the
  mirror-image reflection symmetry in addition to the XY symmetry. A heuristic
  field-theoretical model of the transition has a decoupled fixed point in the
  3D XY universality, and our Monte Carlo simulation suggests that there is
  such a temperature region where long-wavelength fluctuations can be
  described by this fixed point. However, it is shown using scaling arguments
  that the decoupled fixed point is unstable against a fluctuation-induced
  biquadratic interaction, indicating that a crossover to nontrivial critical
  phenomena with different exponents appears as one approaches the critical
  point beyond the transient temperature region. This new scenario clearly
  contradicts the previous notion of the 3D XY universality.
}

\kword{
  interlayer frustration; 
  finite-temperature phase transition; 
  order by disorder; 
  Z$_2$ symmetry breaking;
  BaCuSi$_2$O$_6$
}

\begin{document}
  \maketitle

  \section{Introduction}
  \label{sec:introduction}
  Field-induced critical phenomena of gapped spin-dimer antiferromagnets have
  drawn much attention. Such antiferromagnets typically consist of strongly
  coupled spin-$1/2$ dimers, and are essentially in singlet states in zero
  field. Elementary excitations in the gapped phase under external magnetic
  fields are $S_{z} = 1$ triplet excitations, sometimes called ``triplons,''
  for which a magnetic field acts as a chemical potential. They undergo
  Bose-Einstein condensation (BEC) when their density is appropriately tuned.
  \cite{
    giamarchi-ruegg-tchernyshyov-2008
  }

\begin{figure}
    \centering
    \includegraphics[clip,height=2.75cm,bb=0 0 344 136]{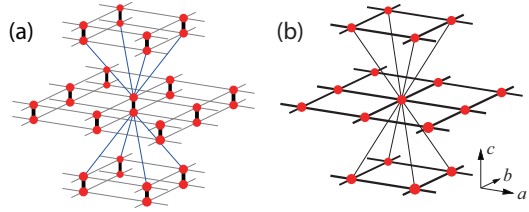}
    \caption{
      (Color online)
      (a) Localized spin degrees of freedom in BaCuSi$_2$O$_6$.
      (b) BCT lattice. 
      For clarity, interlayer bonds are drawn partially.
    }
    \label{fig:bct}
  \end{figure}

  BaCuSi$_2$O$_6$ is one of such spin-dimer compounds,
  \cite{
    sasago-et-al-1997, 
    jaime-et-al-2004, 
    sparta-roth-2004,
    sebastian-et-al-2005, 
    samulon-et-al-2006, 
    sebastian-et-al-2006,
    ruegg-et-al-2007, 
    kramer-et-al-2007, 
    rosch-vojta-2007-1, 
    rosch-vojta-2007-2,
    batista-et-al-2007, 
    schmalian-batista-2008,
    laflorencie-mila-2009
  }
  with characteristic frustration in interlayer interactions. Spin dimers in
  this compound align on the body-centered tetragonal (BCT) lattice
  (Fig.~\ref{fig:bct}). Owing to the lattice geometry, loops that include
  interlayer hoppings are frustrated, which leads to cancellation among
  interlayer interactions. An interesting behavior related to this interlayer
  frustration has been reported.
  \cite{
    comment-on-layer-modulation
  }
  The phase boundary around the quantum critical point is described by the
  power law
  $
  T_{c} \left(H\right) \propto \left(H - H_{c}\right)^{\phi}
  $
  with an anomalous exponent $\phi = 1$.
  \cite{
    sebastian-et-al-2006
  }
  Since mean-field treatment yields $\phi = 2 / d$, the exponent is regarded
  as a characteristic of two-dimensional (2D) systems, and in this sense the
  phenomenon is called ``dimensional reduction.'' We refer to several recent
  papers for further details on this subject.
  \cite{
    sebastian-et-al-2006, 
    rosch-vojta-2007-1, 
    rosch-vojta-2007-2, 
    batista-et-al-2007, 
    schmalian-batista-2008,
    laflorencie-mila-2009
  }
  
  Our main interest in this paper is to study the critical properties at the
  finite temperature transition in BaCuSi$_2$O$_6$. Owing to the broken U(1)
  symmetry of the ordered state and a few experimental observations such as
  the $\lambda$-peak of specific heat, 
  \cite{
    sebastian-et-al-2005
  }
  it has been presumed so far that the transition is in the three-dimensional
  (3D) XY universality class. However, as we will argue in this paper, the
  additional Z$_2$ symmetry breaking that characterizes the ordered phase in
  the BCT lattice makes the XY fixed point unstable. 
  \cite{
    rosch-vojta-2007-1, 
    rosch-vojta-2007-2, 
    batista-et-al-2007,
    schmalian-batista-2008
  }
  The 3D XY-type order takes place in two subsystems or sublattices, namely,
  the even- and odd-numbered layers. The reason for this is that a Z$_2$
  mirror-image reflection symmetry of the BCT lattice precludes any bilinear
  effective coupling between layers on different sublattices. Although this
  Z$_2$ symmetry allows biquadratic interlayer couplings making the order
  ``collinear,'' whether the XY antiferromagnetic (AF) moments of the
  sublattices are parallel or antiparallel remains undetermined, and one of
  them is selected via spontaneous Z$_2$ symmetry breaking. The structure of
  this symmetry breaking is clearly different from that for the standard XY
  ordering, and  our final goal is to understand the effect of the additional
  Z$_2$ symmetry breaking on the universality class of the transition. For
  this purpose, we will study a classical spin model that will be introduced
  in the next section. A Hamiltonian for classical spins is adequate for
  describing the critical behavior near the critical temperature because the
  relevant (largest) fluctuations are classical: fluctuations in the imaginary
  time direction become negligible  because they are confined to a finite size
  $\beta = 1/T$. We will also present numerical simulation results that
  elucidate the above-mentioned symmetry-breaking structure.

  \section{Model}
  \label{sec:model}
  The low-energy subspace generated by the $S_{z} = 1$ triplet and singlet
  well approximates the spin-dimer systems in an applied magnetic field.
  \cite{
    giamarchi-ruegg-tchernyshyov-2008, 
    giamarchi-tsvelik-1999
  }
  The corresponding effective Hamiltonian is the XXZ model for $S=1/2$
  pseudo-spins that represent the two states of each dimer: $S^z=1/2$ for the
  triplet and $S^z=-1/2$ for the singlet. Since the the thermodynamic phase
  transition is driven by classical (or thermal) fluctuations, we can replace
  $S=1/2$ pseudo-spins with classical spins, in order to study the critical
  phenomena near the finite-temperature transition:
  \begin{equation}
    \mathcal{H}_{\textit{cl}} 
    = J_{\parallel} \sum_{n, \left\langle \bm{r},\bm{r}'\right\rangle} 
    \bm{S}_{n, \bm{r}} \cdot \bm{S}_{n, \bm{r}'}
    + J_{\perp} \sum_{n, \bm{r}, \delta\bm{r}} 
    \bm{S}_{n, \bm{r}} \cdot \bm{S}_{n + 1, \bm{r} + \delta \bm{r}}\\
    - H \sum_{n, \bm{r}} S_{n, \bm{r}}^{z},
    \label{eq:Hamiltonian}
  \end{equation}
  where $\bm{S}_{n, \bm{r}}$ is a three-component classical spin, located at a site
  $\bm{r}$ in the $n$-th layer on the BCT lattice ($\bm{r}$ refers to a two-component
  vector),
  $
  \left\langle \bm{r}, \bm{r}'\right\rangle
  $
  are nearest-neighbor pairs on a given layer, and
  $
  \delta\bm{r}=\left(\pm a/2, \pm a/2\right)
  $
  are interlayer displacement vectors (hereafter, we take $a = 1$).
  $J_{\parallel}$ ($J_{\perp}$) is the AF intralayer (interlayer) interaction.
  Throughout the paper, we consider the case 
  $J_{\parallel} > J_{\perp} > 0$,
  whose inequality sign is the same as that in the relation between the
  magnitudes of interdimer exchanges in BaCuSi$_2$O$_6$.
  The finite magnetic field $H$ breaks O(3) spin symmetry down to O(2)
  symmetry, and thus the spins we treat are XY-like. This justifies the
  neglect of the easy-plane-type spin-anisotropy that exists in the effective
  XXZ model mentioned above. Although the two-component spins would serve the
  present purpose just as well as the three-component spins, we use the latter
  for technical reasons.
  
  First, we will discuss the basic properties of $\mathcal{H}_{\textit{cl}}$.
  The energy is minimized when the spins form a canted AF order in each layer.
  The ground-state configuration is given by
  \begin{equation}
    \bm{S}_{n, \bm{r}} = 
    \bigl(
    \sin\Theta\cos\Phi_{n}\,\textrm{e}^{i \bm{Q} \cdot (\bm{r} - \bm{r}_{0}^{\left(n\right)})},\,
    \sin\Theta\sin\Phi_{n}\,\textrm{e}^{i \bm{Q} \cdot (\bm{r} - \bm{r}_{0}^{\left(n\right)})},\,
    \cos\Theta
    \bigr),
  \end{equation}
  with
  $\cos\Theta = H / \left[8\left(J_{\parallel} + J_{\perp}\right)\right]$,
  $\bm{Q} = \left(\pi, \pi\right)$,
  and 
  $\bm{r}_{0}^{\left(n\right)} \equiv \left(\frac{1}{2},
  \frac{1}{2}\right)\delta_{\left(-1\right)^n, 1}$. 
  We define
  $\bm{M}_{XY}^{(n)} \equiv \left(\cos\Phi_{n}, \sin\Phi_{n}\right)$
  to represent the XY AF moment of the $n$-th layer, and in what follows we
  use the term ``AF moment'' to refer to this quantity unless otherwise
  specified. Interlayer mean-field interactions cancel out in the ground state
  because of a combination of the intralayer AF order and the geometry of the
  BCT lattice. This means that the AF moments of one layer can be rotated
  without changing ground-state energy. Thus, the system may be viewed as a
  set of independent 2D layers at $T = 0$.
  
  This apparent 2D character is lifted by thermal fluctuations. We will show
  that these fluctuations select a state qualitatively analogous to the
  ordered phase of the original quantum system. One of the simplest ways to
  see how this ``order by disorder'' 
  \cite{
    villain-et-al-1980, 
    henley-1989
  }
  takes place is to use spin-wave approximation and evaluate free energy as a
  function of the ground-state configuration $\left\{\Phi_{n}\right\}$. Let
  $\theta_{n, \bm{r}}$ and $\phi_{n, \bm{r}}$ represent small fluctuations
  around a given ground-state configuration. Expanding the Hamiltonian to the
  second order in these variables, we rewrite it in the form
  $
  \mathcal{H}_{\textit{cl}} \approx \mathcal{H}_{\textit{sw}} 
  \left(\Theta, \left\{\Phi_{n}\right\}; \{\theta_{n,\bm{r}}\}, \{\phi_{n, \bm{r}}\}\right) 
  = \mathcal{H}_{\text{2D}} + gV
  $
  with $g \equiv J_{\perp} / J_{\parallel}$. Here,
  \begin{equation}
    \mathcal{H}_{\text{2D}} 
    = \frac{1}{2} \sum_{n, \bm{q}}\left(
    \omega_{\theta}\left(\bm{q}\right)\theta_{n, \bm{q}}\theta_{n, -\bm{q}}
    + \omega_{\phi}\left(\bm{q}\right)\phi_{n, \bm{q}}\phi_{n, -\bm{q}}
    \right)
  \end{equation}
  and
  \begin{equation}
    V=\sum_{n, \bm{q}} 
    \Bigl[
      \gamma^{(n)}_{\theta\theta}\left(\bm{q}\right)
      \theta_{n, \bm{q}}\theta_{n+1, -\bm{q}}
      + \gamma^{(n)}_{\phi\phi}\left(\bm{q}\right)
      \phi_{n, \bm{q}}\phi_{n+1, -\bm{q}}
      + \gamma^{(n)}_{\theta\phi}\left(\bm{q}\right)
      \left(
      \theta_{n, \bm{q}}\phi_{n+1, -\bm{q}}
      - \theta_{n+1, -\bm{q}}\phi_{n, \bm{q}}
      \right)
      \Bigr],
  \end{equation}
  where
  $
  \theta_{n, \bm{q}} 
  = {(L^2)}^{-1/2}\sum_{\bm{r}}\theta_{n, \bm{r}}\,\textrm{e}^{-i \bm{q}\cdot\bm{r}}
  $
  and
  $
  \phi_{n, \bm{q}} 
  = {(L^2)}^{-1/2}\sum_{\bm{r}}\phi_{n, \bm{r}}\,\textrm{e}^{-i \bm{q} \cdot\bm{r}}
  $
  with $L^2$ being the number of sites in each layer. 
  The coefficients in $\mathcal{H}_{\text{2D}}$ are written as
  \begin{align}
    \omega_{\theta}\left(\bm{q}\right) 
    &= 2J_{\parallel}
    \left[2 + \left(1 - 2\cos^{2}\Theta\right)\left(\cos q_{x} + \cos
      q_{y}\right)
      \right]
    \\
    \omega_{\phi}\left(\bm{q}\right)
    &= 2J_{\parallel}\sin^2\Theta\left(2 - \cos q_{x} - \cos q_{y}\right),
  \end{align}
  and those in $V$ are written as
  \begin{align}
    \gamma^{(n)}_{\theta\theta}\left(\bm{q}\right)
    & = 4J_{\parallel}\left[
      \sin^2\Theta\cos\frac{q_{x}}{2}\cos\frac{q_{y}}{2}
      - \cos^2\Theta\cos\left(\Phi_{n+1} - \Phi_{n}\right)\sin\frac{q_{x}}{2}\sin\frac{q_{y}}{2}
      \right]
    \\
    & = C_{1}\left(\bm{q}\right) + C_{2}\left(\bm{q}\right)\cos\left(\Phi_{n+1} - \Phi_{n}\right)
    \notag
    \\
    \gamma^{(n)}_{\phi\phi}\left(\bm{q}\right)
    &= -4J_{\parallel}
    \sin^2\Theta\cos\left(\Phi_{n+1} - \Phi_{n}\right)\sin\frac{q_{x}}{2}\sin\frac{q_{y}}{2}
    \\
    \gamma^{(n)}_{\theta\phi}\left(\bm{q}\right)
    &= 4J_{\parallel}
    \sin\Theta\cos\Theta\sin\left(\Phi_{n+1} - \Phi_{n}\right)\sin\frac{q_{x}}{2}\sin\frac{q_{y}}{2},
  \end{align}
  where 
  $
  C_{1} \equiv
  4J_{\parallel}\sin^2\Theta\cos\frac{q_{x}}{2}\cos\frac{q_{y}}{2}
  $ 
  and
  $
  C_{2} \equiv
  -4J_{\parallel}\cos^2\Theta\sin\frac{q_{x}}{2}\sin\frac{q_{y}}{2}
  $.
  Then we expand free energy as
  $
  F = -T\ln Z_{0} 
  - T\sum_{k=1}^{\infty}\frac{\left(-g\right)^k}{k!}
  \beta^{k}\langle V^{k}\rangle_{0}^{(c)}
  $ 
  with $g$ being a small parameter
  ($Z_{0} \equiv \mathrm{Tr}\,\textrm{e}^{-\beta\mathcal{H}_{\text{2D}}}$ and
  $\langle V^{k}\rangle_{0}^{(c)}$ denote the cumulants with respect to
  $Z_{0}^{-1}\textrm{e}^{-\beta\mathcal{H}_{\text{2D}}}$).
  To the fourth order in $g$, we obtain
  \begin{equation}
    F = -T\ln Z_{0}
    - \frac{g^2}{2!}A\left(\Theta\right)TL^2\sum_{n}\cos^{2}\left(\Phi_{n+1} - \Phi_{n}\right)
    - \frac{g^4}{4!}B\left(\Theta\right)TL^2\sum_{n}\cos\left(\Phi_{n+2} - \Phi_{n}\right)
    - \dots,
    \label{eq:free-energy}
  \end{equation}
  with the definitions of $A$ and $B$ given below. We have dropped several
  ``biquadratic'' terms of $O(g^4)$ because they do not change the $O(g^2)$
  term's symmetry discussed below.
  
  The $\left\{\Phi_{n}\right\}$-dependence of $F$ lifts part of the
  ground-state degeneracy, which is unrelated to the system symmetry. The
  coefficient $A\left(\Theta\right)$ of the $O(g^2)$ term is determined by the
  integral
  \begin{align}
    \beta^2{\langle V^2\rangle}_{0}^{(c)}
    &= L^2\sum_{n}\int_{\textrm{BZ}}\frac{d^2 q}{(2\pi)^2}
    \left[
      \frac{{\gamma^{(n)}_{\theta\theta}(\bm{q})}^2}
	   {{\omega_{\theta}(\bm{q})}^2}
	   + \frac{{\gamma^{(n)}_{\phi\phi}(\bm{q})}^2}
	   {{\omega_{\phi}(\bm{q})}^2}
	   + 2\frac{{\gamma^{(n)}_{\theta\phi}(\bm{q})}^2}
	   {{\omega_{\theta}(\bm{q})}{\omega_{\phi}(\bm{q})}}
	   \right]
    \notag
    \\
    &= A\left(\Theta\right)L^2\sum_{n}\cos^{2}\left(\Phi_{n+1} - \Phi_{n}\right) 
    + \textit{const.}
    \label{eq:A}
  \end{align}
  $A$ is found to be positive, favoring collinear configurations where
  $\Phi_{n+1} - \Phi_{n} = 0$ or $\pi$.
  Therefore, this term represents the biquadratic effective interaction
  between nearest-neighbor layers. On the other hand, there is no terms
  proportional to
  $\cos\left(\Phi_{n+1} - \Phi_{n}\right)$ in $F$,
  being consistent with the fact that bilinear effective interactions 
  between nearest-neighbor layers are forbidden by the symmetry of the BCT lattice.
  \cite{
    rosch-vojta-2007-1, 
    rosch-vojta-2007-2, 
    batista-et-al-2007, 
    schmalian-batista-2008
  } 
  A mirror-image transformation with respect to the (100) plane that contains
  $(0, 0)$ or $(1/2, 1/2)$ is such a symmetry operation.
  By the mirror-image transformation with respect to the plane that contains
  $(0, 0)$, for example, sites $\bm{r}$ with 
  $\textrm{e}^{i \bm{Q} \cdot (\bm{r} - \bm{r}_{0}^{\left(n\right)})} = \pm 1$
  in even-numbered (odd-numbered) layers are mapped in sites $\bar{\bm{r}}$ on
  the same layer with $\textrm{e}^{i \bm{Q} \cdot (\bar{\bm{r}} -
  \bm{r}_{0}^{\left(n\right)})} = \pm 1$ ($\textrm{e}^{i \bm{Q} \cdot
  (\bar{\bm{r}} - \bm{r}_{0}^{\left(n\right)})} = \mp 1$).
  Consequently, under the corresponding symmetry transformation 
  $
  \bm{S}_{n, \bm{r}}\to{\bm{S}'}_{n, \bm{r}}\equiv\bm{S}_{n, \bar{\bm{r}}}
  $,
  the local AF moments change as
  $
  \textrm{e}^{i \bm{Q} \cdot (\bm{r} - \bm{r}_{0}^{\left(n\right)})}
  {S}^{a}_{n, \bm{r}}
  \to
  \textrm{e}^{i \bm{Q} \cdot (\bm{r} - \bm{r}_{0}^{\left(n\right)})}
  {S\,'}^{a}_{n, \bm{r}}
  = (-1)^{n} \textrm{e}^{i \bm{Q} \cdot (\bar{\bm{r}} - \bm{r}_{0}^{\left(n\right)})} 
  {S}_{n, \bar{\bm{r}}}^{a}
  $
  ($a = x, y$). 
  This means
  $\Phi_{n}\to\Phi_{n}~\text{($n$: even)}$ and
  $\Phi_{n}\to\Phi_{n}+\pi~\text{($n$: odd)}$ for the phase of their spatial
  average over a layer, resulting in
  $\cos\left(\Phi_{n + 1} - \Phi_{n}\right)$ being mapped to 
  $-\cos\left(\Phi_{n + 1} - \Phi_{n}\right)$.
  Therefore, the two types of the collinear configurations, namely those with
  AF moments of adjacent layers being parallel ($\Phi_{n + 1} - \Phi_{n} = 0$)
  or antiparallel ($\lvert\Phi_{n + 1} - \Phi_{n}\rvert = \pi$), are equivalent.
  While this degeneracy generally exists for the AF moments of
  any two layers $(n, n')$ with $\lvert n - n'\rvert$ being an odd number,
  this is not the case with layers of \textit{even}-numbered separations.
  Indeed, the $O(g^4)$ term in $F$ determined by
  \begin{equation}
    B\left(\Theta\right)		
    \equiv 
    \int_{\text{BZ}}\frac{d^{2}q}{\left(2\pi\right)^{2}}\,
    \frac{
      48\,{C_{1}}^{2}{C_{2}}^{2}
    }{{\omega_{\theta}(\bm{q})}^{4}} > 0
    \label{eq:B}
  \end{equation}
  represents the bilinear effective interaction between second
  nearest-neighbor layers, favoring $\Phi_{n+2} - \Phi_{n} = 0$.
  
\begin{figure}
    \centering
    \includegraphics[clip,height=2.6cm,bb=0 0 350 126]{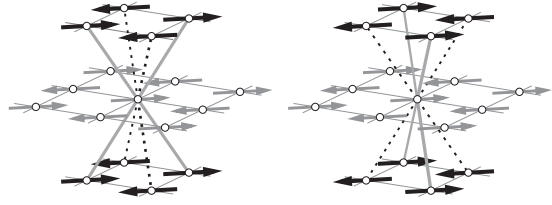}
    \caption{
      Expected ordered phase. 
      Solid and gray arrows denote spins on different sublattices, namely,
      even- and odd-numbered layers, and only their XY components are
      shown for clarity.
      These configurations are distinct in the sense of the spontaneous
      bond ordering (see text).
      Solid (dashed) interlayer lines represent the spin-pairs with parallel
      (antiparallel) XY components.
    }
    \label{fig:bct-ordered-phase}
  \end{figure}

  Although the above spin-wave treatment describes the situation at low
  temperatures $T \ll T_{c}$, we can expect essentially the same form as
  eq. \eqref{eq:free-energy} also for $T \lesssim T_{c}$ in terms of
  symmetry. Therefore, the ordered phase is expected to have the following
  characteristics (see Fig. \ref{fig:bct-ordered-phase}).
  First, there are two subsystems with XY-type 3D ordering,
  namely even- and odd-numbered layers, but there are no bilinear effective
  interactions in between. Second, because of the effective biquadratic
  interactions, the AF moments of these subsystems tend to align in the same
  direction. As a consequence, there are two symmetrically equivalent but
  distinct configurations $\Phi_{n+1} - \Phi_{n} = 0 \text{ or } \pi$
  (Fig. \ref{fig:bct-ordered-phase}). Note that interlayer bonds that are
  equivalent by symmetry become inequivalent in the ordered phase, meaning
  that \textit{bond ordering} results from the spontaneous Z$_2$ symmetry
  breaking. These features are qualitatively the same as the original quantum
  system. 
  \cite{
    rosch-vojta-2007-1, 
    rosch-vojta-2007-2, 
    batista-et-al-2007, 
    schmalian-batista-2008
  } 
  The bond order is a direct 3D analogue of the ``Ising-order'' that
  was discussed in the frustrated square-lattice $J_{1}$-$J_{2}$ model for
  $2J_{2} > J_{1}$.
  \cite{
    henley-1989, 
    chandra-coleman-larkin-1990, 
    loison-simon-2000
  }
  We will use
  \begin{equation}
    \sigma_{n, \bm{r}} = 
    \sum_{\delta\bm{r}}\frac{\left(-1\right)^{\delta {\bm{r}}}}{4}
    \left(
    S^{x}_{n, \bm{r}}S^{x}_{n + 1, \bm{r} + \delta\bm{r}} 
    + S^{y}_{n, \bm{r}}S^{y}_{n + 1, \bm{r} + \delta\bm{r}}
    \right),
    ~\left(-1\right)^{\delta {\bm{r}}}
    \equiv\exp\left[i(\pi, -\pi)\cdot\delta\bm{r}\right]
    \label{eq:sigma}
  \end{equation}
  as the local bond-ordering order parameter.
  It is invariant under O(2) spin rotations but changes its sign
  ($\sigma \to -\sigma$) under mirror-image reflections of
  the lattice, i.e.,
  it serves to detect the Z$_2$ symmetry breaking.  

  \section{Theoretical Arguments on the Phase Transition}
  \label{sec:theoretical arguments on the phase transition}
  
  \subsection{Single bilayer and the BCT lattice}
  \label{subsec:single bilayer and the BCT lattice}
  Let us consider the problem of how many phase transitions take place. There
  are two possible scenarios: 
  a) only one phase transition at $T = T_{c}^{XY} = T_{c}^{BO}$,
  driven by the XY ordering.
  b) two transitions at $T = T_{c}^{BO}$ and $T = T_{c}^{XY}$ ($<
  T_{c}^{BO}$).
  Since the XY ordering necessarily accompanies the bond ordering, 
  $T_{c}^{XY} > T_{c}^{BO}$
  is impossible. We will use the results of a single \textit{bilayer} case to
  argue that a) is the correct scenario. The bilayer case is equivalent to the
  $J_{1}$-$J_{2}$ XY model with $J_{1} = J_{\perp}$ and $J_{2} =
  J_{\parallel}$. This model has been studied numerically by Loison and Simon.
  \cite{
    loison-simon-2000
  }
  It breaks the additional Z$_2$ symmetry for $2J_{2} > J_{1}$ at a finite
  temperature $T = T_{c}^{\kappa}$ with a second-order transition, which is
  followed by a Berezinskii-Kosterlitz-Thouless (BKT)-type transition at a
  slightly lower temperature $T = T_{KT} < T_{c}^{\kappa}$.
  
  We next consider an array of weakly coupled bilayers with $J_{f}$ being the
  inter-bilayer interaction ($J_{f} / J_{\parallel} \ll 1$), such that the
  array returns to the original BCT lattice when $J_{f} = J_{\perp}$. In this
  case, induced by the order-by-disorder mechanism, there appear effective
  interlayer interactions
  $J_{\textit{eff}}'= \left(J_{f}/J_{\parallel}\right)^{k}J_{\parallel}$ 
  between the XY-components of spins and effective interlayer interactions
  $J_{\textit{eff}}'' = \left(J_{f}/J_{\parallel}\right)^{l}J_{\parallel}$
  between bond-ordering order parameters. Equation \eqref{eq:free-energy}
  implies that $k = 4$, and also that $l = 2$ because $J_{\textit{eff}}''$ is
  determined by effective biquadratic interactions. The effective coupling
  $J_{\textit{eff}}'$ is relevant for turning the BKT transition into the true
  long-range XY ordering and $J_{\textit{eff}}''$ drives the 2D bond ordering
  to the 3D behavior. To determine $T_{c}^{XY}$ and $T_{c}^{BO}$, we use a
  simple random phase approximation (RPA) argument.
  \cite{
    janke-matsui-1990
  }
  In this treatment, the XY and bond orderings take place when
  \begin{align}
    J_{\textit{eff}}'\,\chi\left(T_{c}^{XY}\right) \approx 1
    \label{eq:RPA1}
    \\
    J_{\textit{eff}}''\,\chi^{\kappa}\left(T_{c}^{BO}\right) \approx 1
    \label{eq:RPA2}
  \end{align}
  are satisfied, respectively. Here, $\chi\left(T\right)$ and
  $\chi^{\kappa}\left(T\right)$ are the AF XY and 2D bond ordering
  susceptibilities for the single bilayer, respectively. As the temperature is
  lowered, $\chi\left(T\right)$ is expected to diverge exponentially as
  $
  \chi\left(T\right) 
  \propto {J_\parallel}^{-1} \exp\left(b\sqrt{\frac{T_{KT}}{T - T_{KT}}}\right)
  $
  with $b$ being a constant. On the other hand, $\chi^{\kappa}\left(T\right)$ is
  expected to show the power-law divergence
  $
  \chi^{\kappa}\left(T\right) \propto {J_{\parallel}}^{-1}\left(\frac{T -
    T_{c}^{\kappa}}{T_{c}^{\kappa}}\right)^{-\gamma}
  $
  ($\gamma > 0$).
  By substituting these expressions in eqs. \eqref{eq:RPA1} and
  \eqref{eq:RPA2}, we obtain:
  \begin{align}
    \frac{T_{c}^{XY} - T_{KT}}{T_{KT}}
    &\approx
    \left[
      \frac{b}{\ln\left(J_{\textit{eff}}'/J_{\parallel}\right)}
      \right]^2
    = \left[
      \frac{b / k}{\ln\left(J_{f}/J_{\parallel}\right)}
      \right]^2
    \label{eq:Tc_XY}
    \\
    \frac{T_{c}^{BO} - T_{c}^{\kappa}}{T_{c}^{\kappa}}
    &\approx
    \left(
    J_{\textit{eff}}'' / J_{\parallel} 
    \right)^{1/\gamma}
    = \left(
    J_{f} / J_{\parallel}
    \right)^{l/\gamma}.
    \label{eq:Tc_BO}
  \end{align}
  Because eq. \eqref{eq:Tc_XY} diverges as $J_{\textit{f}}$ approaches
  $J_{\parallel}$ while eq. \eqref{eq:Tc_BO} does not, these equations imply
  that $T_{c}^{XY} > T_{c}^{BO}$ for $J_{f} > J_{f}^{c}$. Here, $J_{f}^{c}$
  depends on the difference $T_{c}^{\kappa} - T_{KT}$. Since the difference
  seems to be very small according to the existing numerical simulations,
  \cite{
    how-close-to-each-other
  }
  we can expect that $J_{f}^{c}$ is small. Therefore, we can conclude that the
  thermodynamic phase transition of the XY ordering first takes place as
  temperature decreases over a wide range of $J_{f}$ values. Here, note that
  the above RPA estimate of $T_{c}^{BO}$ is based on assumption that the XY
  spin ordering is absent. Since the XY ordering also breaks the Z$_2$
  symmetry, the bond ordering transition temperature cannot be lower than
  $T_{c}^{XY}$. Therefore, the above RPA result $T_{c}^{XY} > T_{c}^{BO}$
  actually implies a single-phase transition.

  \subsection{Stability of the decoupled XY fixed point}
  \label{subsec:stability of the decoupled xy fixed point}
  The next question is about the universality class of the phase transition,
  in particular as to whether it belongs to the previously expected 3D XY
  universality class. Introducing the ``continuous'' O(2) real vectors
  ${\phi}_{i}^{a}(r)$ ($a = x, y$) to describe the spins on the even- ($i = 1$)
  and odd-numbered ($i = 2$) layers, 
  \cite{
    caution-on-notation-of-phi
  }
  we consider the Landau-Ginzburg-Wilson (LGW)-type effective
  Hamiltonian of the form
  \begin{multline}
    \mathcal{H}_{\textit{eff}} = \int\Bigl[
      \frac{1}{2}\sum_{\mu}
      \left(
      \partial_{\mu}{\phi}_{1}\cdot\partial_{\mu}{\phi}_{1}
      + \partial_{\mu}{\phi}_{2}\cdot\partial_{\mu}{\phi}_{2}
      \right)
      +t
      \left(
      \lvert{\phi}_{1}\rvert^2
      + \lvert{\phi}_{2}\rvert^2
      \right)
      +u
      \left(
      \lvert{\phi}_{1}\rvert^4
      + \lvert{\phi}_{2}\rvert^4
      \right)
      \\
      +\lambda
      \left({\phi}_{1}\cdot{\phi}_{2}\right)^2
      +g
      \lvert{\phi}_{1}\rvert^2 \lvert{\phi}_{2}\rvert^2
      \Bigr]
    d^{d}r.
    \label{eq:H_eff}
  \end{multline}
  The first three terms constitute a standard $\phi^{4}$ theory for the
  decoupled O(2) model. The $\left({\phi}_{1}\cdot{\phi}_{2}\right)^2$ term
  represents a quadrupole-quadrupole interaction induced by the
  order-by-disorder mechanism. The other quartic term
  $\lvert{\phi}_{1}\rvert^2 \lvert{\phi}_{2}\rvert^2$ is included here
  explicitly, because it is generated through renormalization group (RG)
  iterations. As we mentioned earlier, the lattice-symmetry of the original
  model eq. \eqref{eq:Hamiltonian} does not allow effective bilinear
  interactions between nearest-neighbor layers. For this reason, the quadratic
  term ${\phi}_{1}\cdot{\phi}_{2}$ is impossible in eq. \eqref{eq:H_eff}. 
  
  This Hamiltonian eq. \eqref{eq:H_eff} is an $N = M = 2$ case of the model
  referred to as the ``$N$-coupled $M$-vector model,'' 
  \cite{
    domb-green-aharony-1976
  }
  with the additional $\left({\phi}_{1}\cdot{\phi}_{2}\right)^2$ term. It has
  a trivial \textit{decoupled fixed point} (D) at $u \ne 0$ and $\lambda = g =
  0$, which is a plausible candidate for the fixed point corresponding to the
  expected 3D XY universality. This model was first introduced in the 1970s,
  \cite{
    aharony-1975
  }
  in the context of the replica theory for random systems. It was found that
  the decoupled fixed point is \textit{unstable} against perturbations such as
  biquadratic ones. Below, we briefly summarize the argument, because the
  original argument was made in a relatively different context.
  
  The stability of a fixed point against a given perturbation is determined by
  the RG of its conjugate field. Therefore, we need to compute the RG
  eigenvalues $y_{\lambda, D}$ ($y_{g, D}$) of the coupling $\lambda$ ($g$) at
  the decoupled fixed point. They can be computed via two-point correlators at
  the decoupled fixed point, which in this case can be readily factorized into
  known correlators. First, 
  \begin{align}
    \left\langle
    \lvert{\phi}_{1}(r)\rvert^2 \lvert{\phi}_{2}(r)\rvert^2\,\,
    \lvert{\phi}_{1}(r')\rvert^2 \lvert{\phi}_{2}(r')\rvert^2
    \right\rangle_{D}
    &=
    \left\langle
    \lvert{\phi}_{1}(r)\rvert^2 \lvert{\phi}_{1}(r')\rvert^2
    \right\rangle_{D}
    \left\langle
    \lvert{\phi}_{2}(r)\rvert^2 \lvert{\phi}_{2}(r')\rvert^2
    \right\rangle_{D}
    \notag
    \\
    &\propto\lvert r - r'\rvert^{-4x_{t}},
    \label{eq:correlator1}
  \end{align} 
  where $x_{t} = d - 1/\nu$ is the scaling dimension of the energy-density
  field of the 3D XY model with $\nu$ being the correlation-length
  exponent. This means that the scaling dimension of the
  $\lvert{\phi}_{1}\rvert^2 \lvert{\phi}_{2}\rvert^2$ term is  equal to
  $2x_{t}$. Therefore,
  \begin{equation}  
    y_{g, D} = d - 2x_{t} = 2/\nu - d \approx -0.021815 < 0 \,\text{ in }\, d = 3,
  \end{equation}
  where we used $\nu = 0.67155(27)$. 
  \cite{
    campostrini-hasenbusch-pelissetto-rossi-vicari-2001
  } 
  The negative $y_{g, D}$ indicates that the decoupled fixed point is stable
  against the $\lvert{\phi}_{1}\rvert^2 \lvert{\phi}_{2}\rvert^2$
  term. However, this is not the case with the other
  $\left({\phi}_{1}\cdot{\phi}_{2}\right)^2$ term. This term can be decomposed
  into
  \begin{equation}
    \left({\phi}_{1}\cdot{\phi}_{2}\right)^2
    = \frac{1}{2} \left(
    Q_{1}^{xx}Q_{2}^{xx} + Q_{1}^{xy}Q_{2}^{xy} +
    \lvert{\phi}_{1}\rvert^2 \lvert{\phi}_{2}\rvert^2
    \right),
    \label{eq:decomposition}
  \end{equation}
  where
  $Q_{i}^{xx} = (\phi_{i}^{x})^2 - (\phi_{i}^{y})^2$ 
  and 
  $Q_{i}^{xy} = 2\phi_{i}^{x}\phi_{i}^{y}$ ($i = 1,2$) 
  are components of the traceless symmetric tensor of the quadrupole order
  parameter. Using eq. \eqref{eq:decomposition} and the O(2) invariance of
  $\mathcal{H}_{\textit{eff}}$, we obtain
  \begin{multline}  
    4\,\bigl\langle
    \left(\phi_{1}\cdot\phi_{2}\right)^2 (r)\,\,
    \left(\phi_{1}\cdot\phi_{2}\right)^2 (r')
    \bigr\rangle_{D} 
    \\
    = 2\left\langle
    Q_{1}^{xx}(r)Q_{1}^{xx}(r')
    \right\rangle_{D}
    \left\langle
    Q_{2}^{xx}(r)Q_{2}^{xx}(r')
    \right\rangle_{D}
    + \left\langle
    \phi_{1}^2(r)\phi_{1}^2(r')
    \right\rangle_{D}
    \left\langle
    \phi_{2}^2(r)\phi_{2}^2(r')
    \right\rangle_{D}
    \\
    = \frac{C_{QQ}}{\lvert r - r'\rvert^{4x_{Q}}} + \frac{C_{tt}}{\lvert r - r'\rvert^{4x_{t}}},
    \label{eq:correlator2}
  \end{multline}
  where $C_{QQ}$ and $C_{tt}$ are nonzero coefficients and $x_{Q}$ is the
  scaling dimension of the quadrupole order parameter. Comparing 
  $x_{Q} \approx 1.237$
  \cite{
    calabrese-parruccini-2005
  } 
  with 
  $x_{t} \approx 1.5109$, 
  \cite{
    campostrini-hasenbusch-pelissetto-rossi-vicari-2001
  } 
  we find that the quadrupole-quadruple correlator gives the most relevant
  contribution to eq. \eqref{eq:correlator2}. Consequently, the scaling
  dimension of the $\left({\phi}_{1}\cdot{\phi}_{2}\right)^2$ term is equal to
  $2x_{Q}$ and we obtain
  \begin{equation}
    y_{\lambda, D} = d - 2x_{Q} \approx 0.526.
    \label{eq:y-value}
  \end{equation}
  The positive $y_{\lambda, D}$ indicates that $\lambda$ is a relevant
  coupling for the decoupled 3D XY fixed point. In other words, the decoupled
  fixed point is unstable under such perturbation.

  \section{Results of the Monte Carlo Simulation}
  \label{sec:monte carlo results}
  In this section, we present the results our Monte Carlo (MC) simulation for the
  Hamiltonian eq. \eqref{eq:Hamiltonian}. All the results shown in what
  follows are obtained for $J_{\perp} / J_{\parallel} = 0.75$ and $H /
  J_{\parallel} = 10.0$. The system size is $L\times L\times \left(L/4\right)$
  with $L = 16, 24, 32, 40$, and $48$. Our main motivation for these
  parameters and the anisotropic aspect ratio is to realize the proper
  configuration in finite-size systems. These considerations are necessary,
  because fluctuation-induced effective couplings, if normalized per site, are
  typically smaller than $J_{\parallel}$ by 2--3 orders of magnitude.
  
\begin{figure}
    \begin{center}
      \includegraphics[clip, width=14cm,bb=50 50 1490 554]{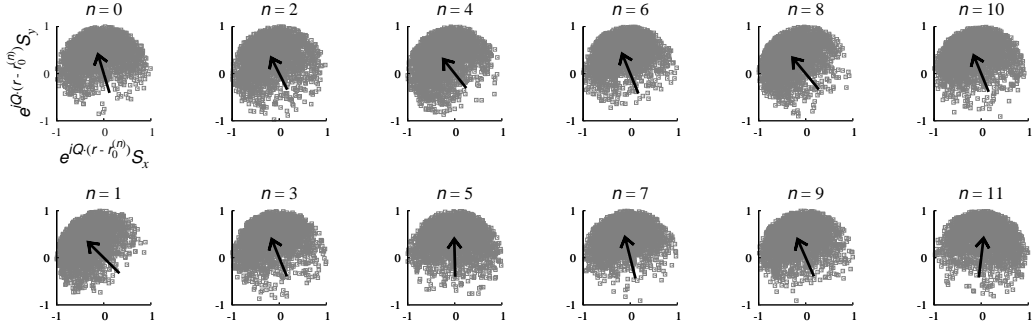}
    \end{center}
    \caption{
      Low-temperature snapshot of a MC simulation for $L = 48$ at $T / J_{\parallel} =
      0.27$ (see text).
    }
    \label{fig:snapshots1}
  \end{figure}

\begin{figure}
    \begin{center}
      \includegraphics[clip, width=14cm,bb=50 50 1490 554]{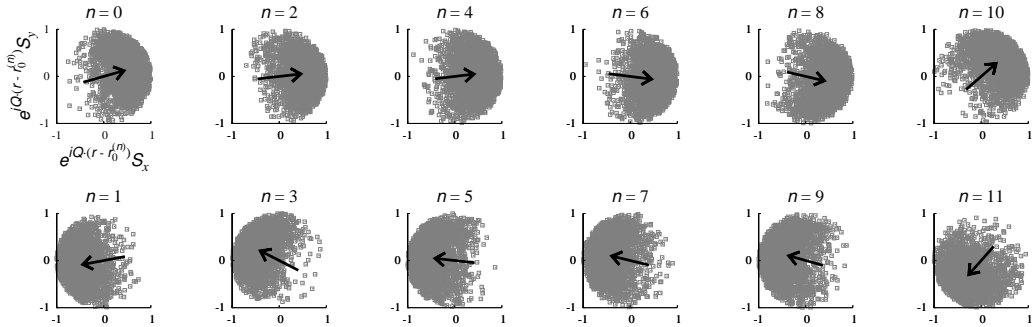}
    \end{center}
    \caption{
      Snapshot under the same conditions as the snapshot in Fig. \ref{fig:snapshots1},
      but obtained using a different random-number sequence.
    }
    \label{fig:snapshots2}
  \end{figure}

  Figures \ref{fig:snapshots1} and \ref{fig:snapshots2} show low-temperature
  snapshots of MC simulations, which are useful for understanding the
  situation. The points in these figures represent the local XY AF moments
  $\textrm{e}^{i \bm{Q} \cdot(\bm{r} - \bm{r}_{0}^{(n)})} (S^x_n,S^y_n)$ in
  each layer, with their spatial positions being discarded for clarity; the
  arrows represent the in-layer AF moments. We can see that the configurations
  are collinear, and that the AF-moments in every other layer tend to align in
  the same direction. By comparing these two figures, we can also see that the
  effective interaction between layers of odd-number separations is not
  bilinear but biquadratic. As we will quantitatively show below, these
  configurations are typical at low temperatures.
  
\begin{figure}
    \begin{center}
      \includegraphics[clip, width=14cm,bb=50 45 770 302]{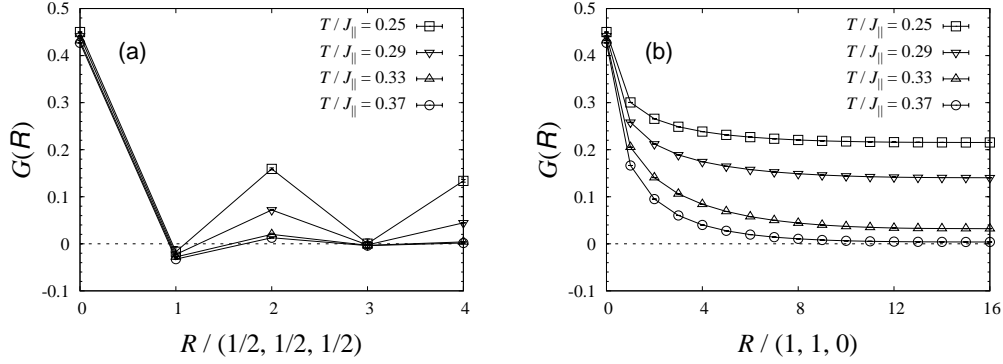}
    \end{center}
    \caption{
      Correlation function $G(\bm{R})$ along [111] (a) and [110] (b) for $L = 32$.
      The lines are guides to the eyes.
    }
    \label{fig:Gr}
  \end{figure}

  In Figs. \ref{fig:Gr}(a) and \ref{fig:Gr}(b), we show decays of the
  correlation function
  $
  G(\bm{R})\equiv
  G\left(\bm{R}(n, \bm{r}; n', \bm{r}')\right) = \bigl\langle 
  S^{x}_{n, \bm{r}} S^{x}_{n', \bm{r}'} + S^{y}_{n, \bm{r}} S^{y}_{n', \bm{r}'}
  \bigr\rangle
  $
  along the interlayer [111] and intralayer [110] directions. $G(\bm{R})$
  along [110] indicates the formation of the in-layer AF order, and positive
  correlations along [111] for $\bm{R} / (1/2, 1/2, 1/2) = 2 \text{ and } 4$
  suggest the presence of a bilinear effective interlayer interaction of the
  ferromagnetic type. On the other hand, the suppression of $G(\bm{R})$ for
  $\bm{R} / (1/2, 1/2, 1/2) = 1 \text{ and } 3$ is mainly due to the
  cancellation of the contributions of opposite (``parallel'' and
  ``antiparallel'') configurations, suggesting the absence of a bilinear
  effective interaction between nearest-neighbor layers; nonzero values at
  these distances are due to short-wavelength fluctuations.
  
  Next we turn to the analysis of the order parameters. There are two relevant
  order parameters to be examined: the XY and bond-ordering order
  parameters. We use
  \begin{equation}
    \bm{M}_{XY} \equiv \frac{8}{L^3} \sum_{n\in\text{even},\, \bm{r}} 
    \textrm{e}^{i \bm{Q} \cdot(\bm{r} - \bm{r}_{0}^{(n)})} 
    \left(S^{x}_{n, \bm{r}}\hat{\bm{x}} + S^{y}_{n, \bm{r}}\hat{\bm{y}}\right),
  \end{equation}
  as the XY order parameter, namely, an AF order parameter defined on the
  ``even'' sublattice (the choice of either ``even'' or ``odd'' is
  arbitrary). On the other hand, we use
  $M_{BO} \equiv \frac{4}{L^3} \sum_{n, \bm{r}} \sigma_{n, \bm{r}}$
  as the bond-ordering order parameter ($\sigma_{n, \bm{r}}$ is defined by
  eq. \eqref{eq:sigma}).
  
\begin{figure}
    \begin{center}
      \includegraphics[clip, width=7cm,bb=50 55 382 549]{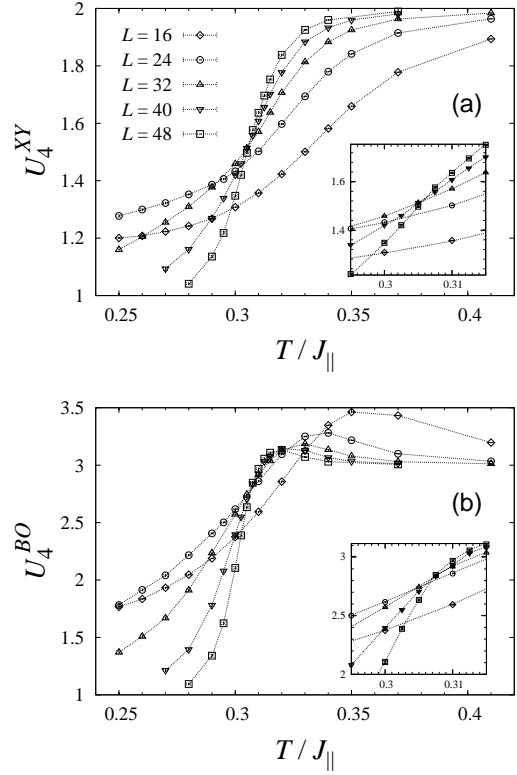}
    \end{center}
    \caption{
      Temperature dependence of the Binder parameters of the
      (a) XY ordering and
      (b) bond ordering.
      The insets show enlarged views in the critical region.
      The lines are guides to the eyes.
    }
    \label{fig:binder-parameters}
  \end{figure}

  Figures \ref{fig:binder-parameters}(a) and \ref{fig:binder-parameters}(b)
  show the temperature dependence of the Binder parameters
  $
  U_4^{XY} \equiv \bigl\langle
  \bm{M}_{XY}^4
  \bigr\rangle 
  / \bigl\langle 
  \bm{M}_{XY}^2
  \bigr\rangle ^2
  $
  and
  $
  U_4^{BO} \equiv \bigl\langle
  M_{BO}^4
  \bigr\rangle 
  / \bigl\langle 
  M_{BO}^2
  \bigr\rangle ^2
  $.
  They should asymptotically show crossings for different system sizes at
  critical points, whereas
  $U_4^{XY} \to 2$ and $U_4^{BO} \to 3$ for $T / J_{\parallel} \gg 1$, 
  and 
  $U_4^{XY}, U_4^{BO} \to 1$ for $T / J_{\parallel} \ll 1$.
  It is not easy to determine $T_{c}^{XY}$ and $T_{c}^{BO}$ precisely as they
  still suffer from severe finite-size effects, but it is clear that there is
  a phase transition. Both $T_{c}^{XY}$ and $T_{c}^{BO}$ are located in the
  region $0.305 < T / J_{\parallel} < 0.310$.
  
\begin{figure}
    \begin{center}
      \includegraphics[clip, width=7cm,bb=50 55 382 549]{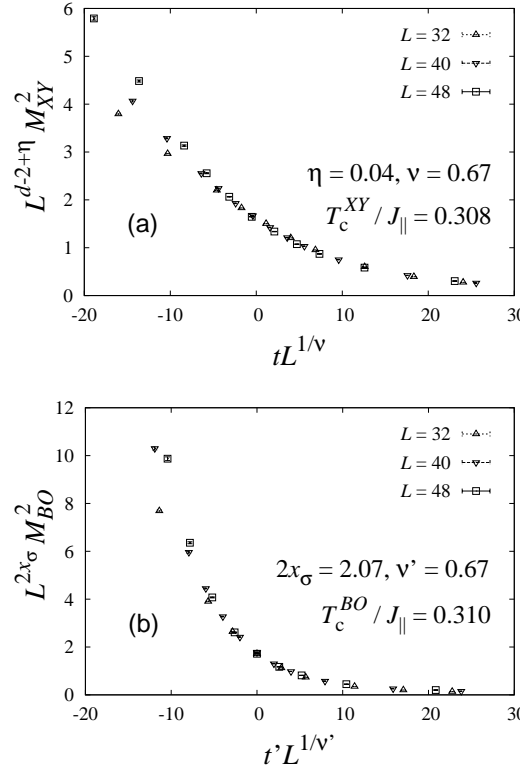}
    \end{center}
    \caption{
      Finite-size scaling plots of
      (a) $\langle\bm{M}_{XY}^2\rangle$ 
      and 
      (b) $\langle M_{BO}^2\rangle$.
    }
    \label{fig:finite-size scaling}
  \end{figure}

  In order to obtain the critical exponents, we perform finite-size
  scaling analysis of the squared quantities 
  $\langle\bm{M}_{XY}^2\rangle$ 
  and 
  $\langle M_{BO}^2\rangle$
  assuming the following standard scaling forms:
  \begin{align}
    \langle\bm{M}_{XY}^2\rangle 
    &= L^{-(d - 2 + \eta)}\,\,\Phi_{XY}\bigl(
    t L^{1/\nu}
    \bigr),
    \\
    \langle M_{BO}^2\rangle 
    &= L^{-2x_{\sigma}}\,\,\Phi_{BO}\bigl(
    t' L^{1/\nu'}
    \bigr),
  \end{align}
  where
  $t \equiv \left(T - T_{c}^{XY}\right)/T_{c}^{XY}$ 
  and 
  $t' \equiv \left(T - T_{c}^{BO}\right)/T_{c}^{BO}$ 
  are reduced temperatures, 
  $\Phi_{XY}$ and $\Phi_{BO}$ are scaling functions,
  $x_{\sigma}$ is the scaling dimension of the bond order parameter,
  and the exponents $\nu$, $\nu'$, and $\eta$ are conventional parameters.
  As shown in Figs. \ref{fig:finite-size scaling}(a) and \ref{fig:finite-size
  scaling}(b), we can produce a reasonable data collapse, where we use
  \begin{equation}
    \eta = 0.04(1),~\nu = 0.67(1), ~\text{and} ~T_{c}^{XY} / J_{\parallel} = 0.308(1)
  \end{equation}
  for the XY ordering and
  \begin{equation}
    2x_{\sigma} = 2.07(1),~\nu' = 0.67(1), ~\text{and} ~T_{c}^{BO} / J_{\parallel} = 0.310(2)
  \end{equation}
  for the bond ordering. These values of $T_{c}^{XY}$ and $T_{c}^{BO}$ are
  consistent with the estimations made from the Binder parameters.
  
  Although there is a slight difference between $T_{c}^{XY}$ and $T_{c}^{BO}$,
  the close proximity of $\nu$ and $\nu'$ (no difference in our resolution)
  suggests that the difference found for $L \le 48$ is due to finite-size
  effects and that there is only one transition. As for the critical
  exponents, their values agree with those of the decoupled 3D XY fixed
  point, for which $\eta_{D}$ and $\nu_{D}$ coincide with those of the 3D XY
  model 
  ($\eta = 0.0380(4)$ and $\nu = 0.67155(27)$
  \cite{
    campostrini-hasenbusch-pelissetto-rossi-vicari-2001
  }
  ) 
  and, in addition, $x_{\sigma, D} = 2x$ holds with $x$ being the scaling
  dimension of the XY order parameter
  (therefore,
  $
  2x_{\sigma, D} = 2\cdot 2x = 2(d - 2 + \eta) = 2.0760(8)
  $
  ).
  \cite{
    schmalian-batista-2008
  }
  The reason for this is that $\sigma$ corresponds to
  ${\phi}_{1}\cdot{\phi}_{2}$ in the continuous-spin language. Its two-point
  correlator can be factorized as
  \begin{equation}
    \bigl\langle
    \left(\phi_{1}\cdot\phi_{2}\right) (r)\,\,
    \left(\phi_{1}\cdot\phi_{2}\right) (r')
    \bigr\rangle_{D} 
    = \bigl\langle
    \phi_{1}(r)\cdot\phi_{1}(r')
    \bigr\rangle_{D} 
    \bigl\langle
    \phi_{2}(r)\cdot\phi_{2}(r')
    \bigr\rangle_{D}
  \end{equation}
  using the O(2) symmetry of the Hamiltonian eq. \eqref{eq:Hamiltonian} or
  \eqref{eq:H_eff}. Then, the simple counting of powers leads to $x_{\sigma,
  D} = 2x$.
  
  \section{Discussion}
  The MC results apparently contradict the previous observation that the
  decoupled fixed point is unstable, but the reason why critical exponents of
  such an unstable fixed point are obtained can be understood through the
  theory of crossover behavior. First, the above MC results suggest that the
  actual RG flow passes through the vicinity of the decoupled fixed point. In
  this parameter region, the system-size dependence of the singular part
  of free energy has the scaling form
  \begin{equation}
    f_{s}\left(t, \lambda, L\right) 
    = \lvert t\rvert^{d\nu_{D}} 
    \Psi\left(tL^{1/\nu_{D}}, \lambda L^{y_{\lambda, D}}\right).
  \end{equation}
  The amplitude of the relevant biquadratic coupling $\lambda$ can be estimated as
  $\lvert\lambda\rvert \sim \left(J_{\perp} / J_{\parallel}\right)^2 A \approx 0.022$
  using eq. \eqref{eq:free-energy}. It is too small to observe any significant
  sign of the crossover behavior in the sense that we need a system as large
  as $L \sim \lvert\lambda\rvert^{-1/y_{\lambda, D}} \approx 1400$, which is
  much larger than the largest system investigated in the present work. In
  this way, the most natural interpretation of our MC results is summarized as
  follows: 
  what we obtained are ``effective'' exponents due to transient behavior of
  the RG flow around the decoupled fixed point. However, since the flow should
  eventually leave the vicinity of the decoupled fixed point, we expect a
  crossover to be observed in larger systems.
  
  The above finite-size scaling arguments can be immediately translated into
  general scaling arguments, in which experimental relevance becomes more
  transparent. The fact that the RG flow passes through the vicinity of the
  decoupled fixed point means that there is a certain temperature region where
  long-wavelength fluctuations are described by the decoupled 3D XY
  model. However, because this fixed point is unstable, this temperature
  region must be outside of the ultimate critical region:
  $\lvert t\rvert \gtrsim t_{X}$ such that the correlation length is bounded as
  $\xi \lesssim \lvert\lambda\rvert^{-1/y_{\lambda, D}}$.
  \cite{
    comment-on-realistic-estimation-of-tX
  }
  We expect that, as one approaches the critical point beyond this
  intermediate temperature region, a crossover from the decoupled 3D XY
  behavior appears.
  
  A question that now arises concerns the nature of the stable fixed
  point. Since the perturbation field $\lambda$ is only slightly relevant at
  the decoupled fixed point for $d$ sufficiently close to $4$ 
  ($y_{\lambda, D} = \frac{3}{5}\epsilon + O(\epsilon^2)$ with 
  $\epsilon \equiv 4 - d$),
  it does not seem to be difficult to find a new stable fixed point by
  performing the $\epsilon$-expansion in the lowest nontrivial order. However,
  the attempt to conduct $O(\epsilon)$ calculation is unsatisfactory.
  \cite{
    aharony-1975
  }
  According to such a calculation, there is no stable fixed point in the
  finite region of the parameter space. Moreover, it fails to reproduce the
  irrelevance of $g$ around the decoupled fixed point, implying that the
  topology of the flow itself may not be convincing unless one goes to
  sufficiently high orders in $\epsilon$. Therefore, it remains to be
  clarified whether a stable fixed point exists and, if it does, what type of
  critical behavior is expected from such a fixed point. We expect that direct
  numerical simulations of the effective Hamiltonian eq. \eqref{eq:H_eff} will
  shed light on this problem. Such a simulation will be performed in a future
  study.

  \section{Conclusions}
  \label{sec:conclusions_and_discussion}
  In this paper, we have studied a classical model of the finite-temperature
  transition in BaCuSi$_2$O$_6$. We have demonstrated that thermal
  fluctuations select a particular configuration in a low-temperature ordered
  phase. It is stabilized by a composite of two subsystems, and is realized
  via multiple symmetry breakings, O(2) and Z$_{2}$, with the latter being
  related to the mirror-image reflection symmetry of the underlying
  lattice. The qualitative characteristics of the phase are the same as those
  of the original quantum system. On the basis of the RPA argument, we have
  also argued that there is only one transition at which the XY and bond
  orderings occur simultaneously in contrast to the 2D single-bilayer case. As
  for the critical behavior of the phase transition, we have shown that a
  plausible form of LGW-type effective Hamiltonian has a decoupled fixed point
  that yields the critical exponents of the 3D XY universality class. However,
  scaling arguments reveal that the decoupled fixed point is unstable against
  the perturbation of the quadrupole-quadrupole interaction. Since the
  quadrupole-quadrupole interaction is induced by the order-by-disorder
  effect, this conclusion clearly excludes the decoupled XY fixed point for
  describing the asymptotic critical behavior. On the other hand, MC
  simulation yields a set of exponents that can be attributed to the decoupled
  3D XY universality. As a first point to be noticed from this observation, we
  have indicated that there is actually an intermediate temperature region
  outside the ultimate critical region, where the observed thermodynamic
  properties are related to the decoupled fixed point. We have also presented
  a reasonable argument explaining why no significant deviation has been
  observed (within our system sizes) from the critical exponents of the
  unstable decoupled fixed point, on the basis of the theory of crossover
  behavior.

  \section*{Acknowledgments}
  We thank H. Tsunetsugu, M. Oshikawa, and D. Uzunov for valuable discussions.
  We also thank T. Suzuki for his critical reading of the manuscript.
  The computation in the present work is executed on computers
  at the Supercomputer Center, Institute for Solid State Physics, University
  of Tokyo.
  The present work is financially supported by 
  the MEXT Global COE Program ``the Physical Science Frontier,'' 
  the MEXT Grand-in-Aid for Scientific Research (B) (19340109), 
  the MEXT Grand-in-Aid for Scientific Research on Priority Areas
  ``Novel States of Matter Induced by Frustration'' (19052004), 
  and the Next Generation Supercomputing Project, Nanoscience Program, MEXT,
  Japan.


\begin{thebibliography}{99}
\bibitem{caution-on-notation-of-phi}
    There should be no confusion with the phase of the local AF moment.
\bibitem{comment-on-realistic-estimation-of-tX}
    Even though the finite-temperature transition can be
    essentially described by the classical theory,
    quantum effects must be taken into account for the quantitative estimate
    of $\lambda$ (and $t_{X}$) for real systems such as BaCuSi$_2$O$_6$.
\end{thebibliography}
\end{document}